\documentclass[aps,twocolumn]{revtex4}
\usepackage{amsmath}
\usepackage{graphicx}

\newcommand\bbbc{{\rm I\!C}}

\newcommand\bbbe{{\rm I\!E}} 

\newcommand\HH{\mathcal{H}}
\newcommand\qed{\bf q.e.d.}
\newcommand\mc{\mathcal}
\newcommand\tr{\textrm{Tr}}

\begin{document}

\title{Quantum Mechanics: Bell and Quantum Entropy for the Classroom}
\author{Philipp Pluch}
\affiliation{Department of Statistics, Klagenfurt University}

\begin{abstract}
In this article we are willing to give some first steps to quantum mechanics
and a motivation of quantum mechanics and
its interpretation for undergraduate students not from physics. 
After a short historical review in the development we
discuss philosophical, physical and mathematical interpretation. We define
local realism, locality and hidden variable theory which ends up in the EPR
paradox, a place where questions on completeness and reality comes into
play. The fundamental result of the last century was maybe Bell's that
states that local realism is false if quantum mechanics is true. From this
fact we can obtain the so called Bell inequalities. After a didactic example
of the fact what these inequalities means we describe the key concept of
quantum entanglement motivated here by quantum information theory. Also
classical entropy and von Neuman entropy is discussed.  
\end{abstract}
\maketitle

\subsection*{The Interpretation}
\begin{minipage}[b]{7cm}
  \begin{em}
    Don't try to understand quantum mechanics or you will fall into a blackhole
    and never be heard from again.
  \end{em}
  Richard Feynman
\end{minipage}

\section{Quantum Mechanics and its Interpretation}
During the last century, quantum theory has proved to be a successful
theory, which describes the physical reality of the mesoscopic and microscopic
world. Up to now, no method is known which contradicts the
predictions made by quantum theory. This is remarkable, since
measurements accuracy has increased, and the size of the systems under
consideration has decreased at a fast pace.\\ 
Quantum mechanics was developed with the aim to describe atoms and
to explain the observed spectral lines in a measurement apparatus. During
the development of quantum mechanics  
the fact that quantum theory allows for an accurate description of reality is
obvious from many physical experiments, and has probably never been seriously
disputed. On the other hand, for the interpretation of quantum mechanics,
things could not be more different. Since the theory of quantum mechanics
has been formulated, following question arise:
\begin{quote}
How can we interpret the mathematical formulation of quantum mechanics?
\end{quote}
This question leads to a discussion, in which people with different
philosophical backgrounds 
gave different answers. 
Quantum theory and quantum mechanics do not account for single
measurement outcomes in a deterministic way. One accepted
interpretation of quantum mechanics is the Copenhagen interpretation. The
Copenhagen manifest argued 
that a measurement causes an instantaneous collapse of the wave function which
describes the quantum system, the system is after the collapse random - pure
chaos. \\ 
The most prominent opponent to the Copenhagen interpretation was Albert
Einstein, who had developed a way from instrumentalism of positivism to a
rational realism. Einstein did not believe in the idea of genuine randomness
in nature, the main argument in the Copenhagen interpretation. In his view
quantum mechanics is incomplete and suggested that there had to be
'hidden' variables, responsible for random measurement results. \\
In fact the famous paper 'Can quantum mechanical description of physical
reality be considered complete?', authored by Einstein, Podolsky and
Rosen in 1935, which condensed the
philosophical discussion into a physical argument. They claim that given a
specific experiment,  in which the outcome of a measurement could be known
before the measurement take place, there must exist something in the real
world, an 'element of reality', which determines the measurement outcome.
They postulate, that these elements of reality are local, in the sense
that they belong to a certain point in space-time. This element may only be
influenced by events which are located in the backward light cone of this
point in space-time. Even though these claims sound reasonable and convincing,
these are assumptions about nature, which are nowadays known as
local realism. \\
\subsection{Local Realism}
The principle of locality is that objects at different places cannot have direct
influence on one another. An object is influenced directly only by its immediate
surroundings. Einstein argued:
\vskip 0.5cm
\emph{The following idea characterises the relative independence of objects
  far apart in space (A and B): external influence on A has no direct
  influence on B; this is known as the Principle of Local Action, which is
  used consistently only in field theory. If this axiom was to be completely
  abolished, the idea of the existence of quasi enclosed systems, and thereby
  the postulation of laws which can be checked empirically in the accepted
  sense, would become impossible.} 
\vskip 0.5cm 	
Local realism is the combination of the principle of locality with the
assumption that all objects must objectively have their properties already
before these properties are observed. Einstein argued with the moon: 
\vskip 0.5cm
\emph{The Moon is out there even when no one is observing it.} 
\vskip 0.5cm
Local realism is a significant feature of classical general
relativity and classical Maxwell's theory, but quantum mechanics
rejects this principle. Every theory that, like quantum mechanics, is
compatible with 
violations of Bell's inequalities must abandon local realism. (The vast
majority of physicists believe that experiments have demonstrated such
violations, but some local realists dispute this with the argumentation of
loopholes in the tests.) Different interpretations of quantum
mechanics reject different parts of local realism. 
\\
In interpretations like the
Copenhagen interpretation where the wavefunction is assumed to have no direct
physical interpretation or reality, the many-worlds
interpretation\, and the 
interpretation based on Consistent Histories, it is realism that is 
rejected. The actual definite properties of a physical system 'do not exist'
prior to the measurement and the wavefunction has a restricted interpretation
as nothing more than a mathematical tool used to calculate the probabilities
of experimental outcomes, in agreement with positivism in philosophy as the
only topic that 
science should discuss.  
\\
In the version of the Copenhagen interpretation where the wavefunction is
assumed to have a physical interpretation or reality (the nature of which is
unspecified), the principle of locality is violated during the measurement
process via the  wave function collapse. This is a non-local process because
Born's 
Rule, when applied to the system's wave function, yields a probability density
for all regions of space and time. Measuring the physical system $S$
the probability density vanishes everywhere instantaneously, except where and
when the measured entity is found to exist. This vanishing is a
physical process, and clearly non-local, if
the wave 
function is considered physically real and the probability density converged
to zero at arbitrarily far distances during the duration for the
measurement process. 
\\
Bohm interpretation wants to preserve realism, and it needs to
violate the principle of locality to achieve the required probability
distributions.
It also violates causality which implies
a conflict with special theory of relativity because real,
superluminal signals would have to be propagated. 
\\
Because of the differences in the interpretations, which are 
philosophical except the one of Bohm, physicists
use an language independent of
the interpretation. In this context, only the measurable action at
a distance - a superluminal propagation of real, physical information - would
be considered as a violation of locality by the physicists. Such
phenomena have never been build by experiments and they are not predicted by 
theories. (Possible exception can be Bohm's theory).
\\
Locality is the main axiom of relativistic quantum field theory in connection
with causality. A formalization of locality is if two observables localized
within two distinct 
space-time regions must commute. This interpretation of
'locality' is closely related to the relativistic version in physics.
\\
In the EPR paper the argument is a thought experiment on pairs
of entangled particles. This experiment shows that both
position and momentum of the particles are elements of reality. 
Quantum mechanics does not include states for which position and momentum are
well-defined (because of the measurement) simultaneously. From this point 
EPR conclude that quantum mechanics is incomplete. The problem is a
description of variables, which correspond to the elements of reality. This
variables are the local hidden variables.

\subsection{Non-locality and Hidden Variable Theories}
Einstein never prize the implications of quantum theory, 
despite the undeniable triumph of quantum theory. Einstein's faith was that
quantum  mechanics could be completed by adding various as-yet-undiscovered
variables. For him hidden variables
would let us to a deterministic description of nature.
\\
\vskip 0.5cm
\emph{'God does not play dice'} 
\\
\vskip 0.5cm
The completeness of quantum mechanics  
was attacked by the Einstein-Podolsky-Rosen Gedanken experiment with the
argument   
that there have to be hidden variables in order to
avoid non-local, instantaneous effects at a distance. The 
position-momentum uncertainty relation served as a guideline for their 
argument, although it is most clear to us with the help of Bohm 
employing a pair of spin-$\frac{1}{2}$ particles in a singlet
state.
\subsection{EPR Paradox}
The EPR paradox draws on a phenomenon predicted by quantum mechanics, known as
quantum entanglement, to show that measurements performed on spatially
separated parts of a quantum system can apparently have an instantaneous
influence on one another, now known as nonlocal behaviour. We
illustrate this 
in a simplified version of the EPR thought experiment due to Bohm.
\\ 
Let us consider a source that emits pairs of electrons, with one electron sent
to Alice and the other to Bob.
We arrange our source in such a way that each emitted electron pair
occupies a quantum state called a spin singlet. This is
a quantum superposition of two states, (a) and (b). In state (a),
electron $A$ has spin upward along the $z$-axis ($+z$) and electron $B$ has
spin downward along the $z$-axis ($-z$). In state (b) the electron $A$ has spin
$-z$ and the electron $B$ has spin $+z$. So it is impossible to associate either
electron in the spin singlet with a state of definite spin. The electrons are
thus said to be entangled.  
\\
Alice measures the spin along the $z$-axis and obtains one of two
possible outcomes: $+z$ or $-z$, suppose she gets $+z$. 
So the quantum state of the system $S$ collapses into state (a). The
quantum state determines the probable outcomes of any measurement performed on
the system. In this case, if Bob subsequently measures spin along the
$z$-axis, he will obtain $-z$ with probability $1$. Similarly, if Alice gets $-z$, Bob will get $+z$.
\\
Let us suppose that Alice and Bob now decide to measure spin along the
$x$-axis. According to quantum mechanics, the spin singlet state may equally
well be expressed as a superposition of spin states pointing in the $x$
direction. We call these states (a1) and (b1). In state (a1), Alice's
electron has spin $+x$ and Bob's electron has spin $-x$. In state (b1),
Alice's electron has spin $-x$ and Bob's electron has spin $+x$. If
Alice measures $+x$, the system collapses into (a1), and Bob will get $-x$. If
Alice measures $-x$, the system collapses into (b1), and Bob will get $+x$. 
\\
Because of quantum mechanics the $x$-spin and $z$-spin are incompatible
observables, this means that there is a Heisenberg uncertainty principle
operating between them: a quantum state cannot possess a definite value for
both variables. Let us now suppose Alice measures the $z$-spin and obtains 
$+z$, so that the quantum state collapses into state (a). Now
Bob measures the $x$-spin. According to
quantum mechanics, when the system is in state (a), Bob's $x$-spin measurement
will be with probability $0.5$ $+x$ and with probability
$0.5$ $-x$. Furthermore, it is fundamentally impossible to predict which
outcome will appear until Bob actually performs the measurement. 
\subsection{Completeness and Reality}
We introduce now two main concepts used by Einstein, Podolsky, and Rosen,
which are crucial to their attack on quantum mechanics:
\begin{enumerate} 
  \item The elements of physical reality
  \item The completeness of a physical theory
\end{enumerate}
EPR did not directly address the philosophical meaning of an 'element
of physical reality'. They made a assumption that if the value of
any physical quantity of a system can be predicted with absolute certainty
prior knowledge to performing a measurement or otherwise disturbing it, then
that quantity corresponds to an element of physical reality. We note here that
the converse is not true. 
\\
EPR defined a complete physical theory as one in which every element
of physical reality is accounted for. They showed that
using these two definitions quantum mechanics is not a complete physical
theory. 
\\
Suppose
Alice decides to measure the value of the
$z$-spin. After Alice's measurement the $z$-spin of Bob's electron
is definitely known, so it is an element of physical reality. After 
Alice's measurement the spin of Bob's
electron is an element of physical reality. 
\\
We conclude that a quantum state cannot possess a definite value for both
$x$-spin and $z$-spin. If quantum mechanics is a complete physical theory the
$x$-spin and $z$-spin cannot be elements of reality at
the same time. So Alice's decision whether to perform a
measurement along the $x$-axis or $z$-axis has an instantaneous effect on the
elements of physical reality at Bob's location. This is a violation of the
principle of locality. 
\\
The principle of locality says that physical processes occurring at one
place should have no immediate effect on the elements of reality at another
location. This is a reasonable assumption to make,
as it seems to be a consequence of special relativity:
information cannot be transmitted faster than lightspeed without
violating causality. It is generally believed that any theory which violates
causality would also be internally inconsistent, and thus deeply
unsatisfactory. 
\\
It turns out that quantum mechanics violates the principle of locality without
violating causality. Causality is preserved because there is no way for Alice
to transmit information to Bob by manipulating her measurement
axis. Which axis she uses, she has a probability $0.5$ of obtaining $+$
and probability $0.5$ of obtaining $-$. 
In the quantum mechanic context, it is impossible for her to influence the
results. On the other hand Bob is only able to perform his measurement
once, the 'no
cloning theorem' makes it impossible to make copies
of the electron he receives to perform a classical statistical analysis.
In the measurement he perform
there is a  probability $0.5$ of getting $+$ and 
probability $0.5$ of getting $-$, regardless of whether or not his axis is
aligned with Alice's one.
\\
We now discuss the impossibility of copying
states. Suppose the state of a quantum system $A$ is a qubit, which we wish to
copy, given by
\begin{eqnarray*}
|\psi\rangle_A = a |0\rangle_A + b |1\rangle_A.
\end{eqnarray*}
with unknown $a, b\in\bbbc$. For a copy we use a system $B$ with identical
Hilbert space $\HH$ and initial state 
\begin{eqnarray*}
|\kappa\rangle_B
\end{eqnarray*} 
which must be independent of $|\psi\rangle_A$, of which we have no prior
knowledge. The composite system is described by the tensor product, 
its state is 
\begin{eqnarray*}
|\psi\rangle_A |\kappa\rangle_B.
\end{eqnarray*}
For manipulating the composite system, we can perform
an observation, which implies a wave function collapse. Alternatively, we
could control the 
Hamiltonian of the system and thus the linear time evolution operator
$U(\Delta t)$. We must fix a time interval $\Delta t$, again independent of
$|\psi\rangle_A$. Then $U(\Delta t)$ acts as a copier provided
\begin{eqnarray*}
U(\Delta t) |\psi\rangle_A |\kappa\rangle_B &=& |\psi\rangle_A |\psi\rangle_B\\
&=& (a |0\rangle_A + b |1\rangle_A)(a |0\rangle_B + b |1\rangle_B)\\
&=& a^2 |0\rangle_A |0\rangle_B + a b |0\rangle_A |1\rangle_B + b a
|1\rangle_A |0\rangle_B \\ &&+ b^2 |1\rangle_A |1\rangle_B
\end{eqnarray*}
for all $\psi$. This holds for the basis states as well, so
\begin{eqnarray*}
U(\Delta t) |0\rangle_A |\kappa\rangle_B &=& |0\rangle_A |0\rangle_B\\
U(\Delta t) |1\rangle_A |\kappa\rangle_B &=& |1\rangle_A |1\rangle_B.
\end{eqnarray*}
Then the linearity of $U(\Delta t)$ implies that
\begin{eqnarray*}
U(\Delta t) |\psi\rangle_A |\kappa\rangle_B &=& U(\Delta t) (a |0\rangle_A + b |1\rangle_A)|\kappa\rangle_B\\
&=& a |0\rangle_A |0\rangle_B + b |1\rangle_A |1\rangle_B\\
&\neq&  a^2 |0\rangle_A |0\rangle_B + a b |0\rangle_A |1\rangle_B \\ && + b a
|1\rangle_A |0\rangle_B + b^2 |1\rangle_A |1\rangle_B. 
\end{eqnarray*}
So $U(\Delta t) |\psi\rangle_A |\kappa\rangle_B $ is not equal to
$|\psi\rangle_A |\psi\rangle_B$, as may be verified by plugging in
$a=b=2^{-\frac{1}{2}}$, so $U(\Delta t)$ cannot act as a general copier. $\qed$
\\
However, the principle of locality appeals powerfully to physical intuition,
and Einstein, Podolsky and Rosen were unwilling to abandon it. Einstein
derided the quantum mechanical predictions as 'spooky action at a
distance'.  
\\
In the EPR paper for any description of nature
following two properties are suggested 
\begin{itemize}
\item Anything that
happens here and now can influence the result of a measurement 
elsewhere, but only if enough time has elapsed for a 
signal to get there without travelling faster than lightspeed 
\item The result of any measurement is predetermined. 
\end{itemize}
EPR discussed the consequences of these two conditions on 
observations of quantum particles that had previously interacted 
with one another. The conclusion was that the particles would 
exhibit correlations that lead to contradictions with Heisenberg's
uncertainty principle, so quantum mechanics is incomplete.
\\
EPR paradox motivated Schr\"odinger's 
to introduce entanglement as the characteristic feature of quantum
mechanics. J. Bell (Bell 1964) 
tried to find a way of showing that the notion of hidden variables could remove
the randomness of quantum mechanics. EPR 
paradox 
was  first nothing more than a philosophical debate for many 
physicists. Bell's theorem concluded that it is 
impossible to mimic quantum theory with the help of a set of local hidden 
variables. Consequently any classical imitation of quantum mechanics  
is non-local. But this fact does not imply the existence of
any non-locality  
in quantum theory itself.

\section{Bell's Theorem}
For more than three decades
no prove of the existence of local hidden variables was formulated. No
empirical methods to prove the existence or non existence of hidden variables
were available.
\\
In 1964 J.S. Bell noticed in his work 'On the Einstein-Podolsky-Rosen paradox'
that the existence of local hidden variables implies a certain inequality (the
so called Bell inequality) between measurement outcomes, while quantum
mechanics predicts measurement outcomes which violate this inequality. In
Bell experiments it is possible to check whether the
predictions of quantum mechanics are correct in the sense of local hidden
variables, or whether nature obeys the Bell inequalities, in the sense that
quantum mechanics would predict wrong measurement results: quantum mechanics
would be wrong rather than incomplete. \\
Bell experiments were discussed very deep in literature and have been
performed and they were excellent agreement with the predictions of quantum
mechanics. However, the importance of Bell's experiment is not due to the fact
that quantum mechanics has once more shown to give a precise description
of nature: it shows that the microscopic world is guided by laws which are
inherently non classical: it is not possible to add something to quantum
mechanics which would make it a classical theory. \\
\subsection{Bell's Gedanken Experiment}
Bell considered a setup in which two observers, Alice and Bob, perform
independent measurements on a system $\mc{S}$ prepared in some fixed state.
Each observer has a detector as a measurement apparatus. On each trial,
Alice and Bob can independently choose between various detector
settings. Alice can choose a detector setting $a$ to obtain a measurement
$A(a)$ and Bob can choose a detector setting $b$ to measure $B(b)$. Alice and
Bob collect statistics on their measurements and correlate the results.  
\\
There are two main assumptions in Bell's analysis:
\begin{enumerate}
\item Each  measurement reveals an objective physical property of the system. 
\item A measurement taken by one observer has no effect on the measurement
  taken by the other.
\end{enumerate} 
Bell's inequality and its experimental violation destroyed the idea that
quantum mechanics can be described by a classical theory. However, the insight
that quantum mechanics is a non classical theory did not only destroy hopes,
but also allowed the dawn of a new era in physics, the so called quantum
physics. Physicists started to realise that if quantum physics is non
classical, it might also allow us to do things which are not possible or at
least not feasible in classical world. As an example we name here the SECOQC
project.

\section{Bell Inequalities}
Quantum mechanics predicts phenomena which are
counterintuitive to a classical understanding of our world's nature so much
that respected physicists have questioned how well the model represents actual
physical reality (see Einstein, 1935). Bell inequalities provide means to
test some of the most counterintuitive predictions of quantum mechanics. In
order to understand how Bell inequalities behave we discuss the concepts of
completeness, locality and entanglement. 
Intuitive from our experience we might expect
that physical systems have definite objective properties. However, any quantum
mechanical model of a system cannot simultaneously describe definite values
for all its physical properties, but instead describes a weighted
superposition of states. As a matter of philosophy, one might choose to
believe that all properties of a system always have definite objective values
and thus that the theory of quantum mechanics is incomplete in its inability
to describe them. These definite values could be described if we have
knowledge of some hidden variables (realist
interpretation of quantum mechanics). We can also believe that
quantum mechanics is a complete theory, so that there exists no hidden
variables in our system $S$, and that physical reality is probabilistic rather
than deterministic, known as orthodox
interpretation.
\\ 
Based on intuition from classical mechanics and special
relativity we might expect physical reality to be local, which means that the
result of any measurement performed by an experiment only depends on the
values of physical properties in the immediate space time vicinity of
measurement. If the outcomes of a measurement in two distinct space and
distinct in time locations can be interdependent, then nature is alocal or non
local.  \\
Two systems are called entangled if they are distinct, if they each exhibit
superposition of states in some property $q$, and if knowledge of $q$ for the
other particle is available. As an example of entangled we can consider a neutral pi
meson that decays into a positron and
an electron, which fly apart. The pion has spin
$0$ and the positron and the electron each have a spin of magnitude
$\frac{1}{2}$, by consideration of angular momentum, measurement of the
$z$-components of the positron's and electrons will be opposite. The orthodox
interpretation of quantum mechanics is that neither particle has a definite
$z$-component of spin before measurements take place. Each of it will be in a
superposition of spin states. However, when one is measured and its
$z$-component of spin is analysed, the $z$-component of the other is
immediately assumed a definite direction. This interpretation clearly violates
locality because the two particles can move arbitrarily far apart before
measurement. If one adopts the realist position, then clearly the particles
have definite $z$-components of spin from the moment they come to existence at
the same point in space-time, and entanglement does not contradict
locality. Bell's theorem shows that any local
hidden variable theory is incompatible with quantum mechanics and this
disagreements can be tested experimentally. Results from such experiments
closely follow quantum mechanical predictions and violate the Bell
inequalities. This shows that our nature cannot be local, regardless of
whether the realist or orthodox interpretation is more accurate. 

\subsection{Bell's inequalities}
In quantum mechanics properties of objects are not clear to
verify. They are only well defined if we perform a measurement. Two quantum
particles that are interacting with each other, the possibility of predicting
properties without measurement  
on either side led to the EPR paradox. The postulation of unknown random
variables,  
hidden variables, would restore localism. On the other hand, randomness 
is intrinsic to quantum mechanics.
\\
Bell implemented an experiment that
would prove it properties are well-defined or not, an experiment that would 
give one result if quantum mechanics is correct and
another result if hidden variables are needed. The most important
are Bell's original  
inequality (Bell, 1964), and the Clauser-Horne-Shimony-Holt (CHSH) inequality 
(Clauser, Horne, Shimony and Holt 1969). In Bell's work:

\vskip 0.5cm
\noindent {\it ``Theoretical physicists live in a classical world, looking out
  into  
a quantum-mechanical world. The latter we describe only subjectively, in terms
of  
procedures and results in our classical domain. (...) Now nobody knows just
where  
the boundary between the classical and the quantum domain is situated. (...)
More  
plausible to me is that we will find that there is no boundary. The wave
functions  
would prove to be a provisional or incomplete description of the
quantum-mechanical  
part. It is this possibility, of a homogeneous account of the world, which is
for me  
the chief motivation of the study of the so-called ``hidden variable"
possibility.  
\\
(...) A second motivation is connected with the statistical character of 
quantum-mechanical predictions. Once the incompleteness of the wave function 
description is suspected, it can be conjectured that random statistical 
fluctuations are determined by the extra ``hidden" variables -- ``hidden" 
because at this stage we can only conjecture their existence and certainly 
cannot control them. 
\\
(...) A third motivation is in the peculiar character of some
quantum-mechanical predictions, which seem almost to cry out for a hidden
variable interpretation. This is the famous argument of Einstein, Podolsky and
Rosen. (...) We will find, in fact, that no local deterministic
hidden-variable theory can reproduce all the experimental predictions of
quantum mechanics. This opens the possibility of bringing the question into
the experimental domain, by trying to approximate as well as possible  
the idealized situations in which local hidden variables and quantum mechanics 
cannot agree."}
\\
\vskip 0.5cm
We will discuss the development of the Bell's original
inequality. With the example advocated  by Bohm and Aharonov (1957), the EPR
argument is the 
following. Let us consider a pair of spin one-half particles in a singlet
state, and we place Stern-Gerlach magnets in order to measure selected
components of the spins ${\sigma}_1$ and  
${\sigma}_2$. If the measurement of the component ${\sigma}_1\cdot {\bf a}$, 
with ${\bf a}$ being some unit vector (observable ${\bf a}$), yields $+1$,
then the quantum mechanics says that measurement of the component 
${\sigma}_2\cdot {\bf a}$ must yield $-1$ and vice versa, because 
the two particles are anticorrelated. Easily seen one can predict in
advance the result of measuring any chosen component of ${\sigma}_2$, by
previously measuring the same component of ${\sigma}_1$. 
\\
Now let us construct a classical description of these correlations. 
Suppose that there exist a continuous hidden variable $\lambda$. The
corresponding outcomes of measuring ${\sigma}_1\cdot {\bf a}$ 
and ${\sigma}_2\cdot {\bf b}$ are $A({\bf a},\lambda)=\pm 1$ and $B({\bf
  b},\lambda)=\pm 1$, respectively. The main assumption is that result $B$ for
particle two is independent of the setting ${\bf a}$, nor $A$ on ${\bf b}$, in
other words, we address individual particles locally. Suppose that
$\rho(\lambda)$ is the probability distribution of $\lambda$. If the
quantum-mechanical expectation value of 
the product of the two components ${\sigma}_1\cdot {\bf a}$ and
${\sigma}_2\cdot {\bf b}$ is  

\begin{equation}\label{Bell_1}
\langle \, {\sigma}_1\cdot {\bf a}\, , {\sigma}_2\cdot {\bf b} \,	\rangle 
\,=\, -\, {\bf a} \cdot {\bf b},
\end{equation}
\noindent then the hidden variable model lead to 
\begin{equation}\label{Bell_2}
P({\bf a},{\bf b}) \, = \, \int \, \rho(\lambda) \, A({\bf a},\lambda) 
B({\bf b},\lambda)d\lambda.
\end{equation}
If the hidden variable description has to be correct, then 
result (\ref{Bell_2}) must be equal to Bell's result. 
We introduce now anticorrelation in this scheme: 
$A({\bf a},\lambda)=-B({\bf a},\lambda)$ and (\ref{Bell_2}) is
\begin{eqnarray*}
P({\bf a},{\bf b}) \, = \, -\int \, \rho(\lambda) \, A({\bf
  a},\lambda)  A({\bf b},\lambda)d\lambda.
\end{eqnarray*}
The extension with one more unit vector ${\bf c}$, we get
\begin{eqnarray} \label{Bell_3}
P({\bf a},{\bf b}) \, -\, P({\bf a},{\bf c}) \,&=& \, -\int \, 
\rho(\lambda) \, [A({\bf a},\lambda)A({\bf b},\lambda)\\ &&-A({\bf
  a},\lambda)A({\bf c},\lambda)] d\lambda\cr 
&=& \, \int \, \rho(\lambda) \, A({\bf a},\lambda) 
A({\bf b},\lambda)\\ && \times [A({\bf a},\lambda)A({\bf c},\lambda)\,-\,1] d\lambda.
\end{eqnarray}
If we consider that $A({\bf a},\lambda)=\pm 1$ and $B({\bf
  b},\lambda)=\pm 1$,  
we get (\ref{Bell_3}) as
\begin{eqnarray*}
|P({\bf a},{\bf b}) \, -\, P({\bf a},{\bf c})| \, &\leq& \, \int \,
\rho(\lambda) \,  
[1\,-\,A({\bf b},\lambda)A({\bf c},\lambda)] d\lambda\,\\ &=&\,1\,+\,P({\bf b},{\bf c}).
\end{eqnarray*}
So Bell's original inequality is given by
\begin{equation} \label{Bell_4}
1\,+\,P({\bf b},{\bf c}) \, \geq \, |P({\bf a},{\bf b}) \, -\, P({\bf a},{\bf
  c})|. 
\end{equation}
By performing an experiment that violates this inequality, 
the local hidden variables theories are not correct. In the case of a singlet
state $|\psi\rangle=1/\sqrt{2}\,(|01\rangle-|10\rangle)$, the quantum
mechanical prediction (\ref{Bell_1}) is equal to $-\rm{cos}({\bf a},{\bf b})$,
which violates Bell's inequality (\ref{Bell_4}) for different angles. 
In the case of the CHSH inequality, we can relax the conditions 
$A({\bf a},\lambda)=\pm 1$ and $B({\bf b},\lambda)=\pm 1$ to 
$|A({\bf a},\lambda)|\leq 1$ and $|B({\bf b},\lambda)|\leq 1$. Proceeding as
before, we get 

\begin{equation} \label{CHSH}
|P({\bf a},{\bf b}) \, -\, P({\bf a},{\bf d})| \, + \, 
|P({\bf c},{\bf d}) \, -\, P({\bf c},{\bf b})| \, \leq \, 2.
\end{equation}
The quantum limit of the CHSH inequality is
\begin{equation} \label{CHSH}
|P({\bf a},{\bf b}) \, -\, P({\bf a},{\bf d})| \, + \, 
|P({\bf c},{\bf d}) \, -\, P({\bf c},{\bf b})| \, \leq \, 2\sqrt{2}.
\end{equation}
These inequalities can be tested experimentally
by using  
random counts. Pairs of particles are emitted as a result of a quantum
process, and further  analysed and detected. In practice perfect
anti correlations are difficult to obtain. Moreover, the system is always
coupled to an environment. Although several experiments validate the
quantum-mechanical view, the issue is not conclusively settled.

\subsection{Example for Bell Inequalities}
The following example (Mermin, 1985) illustrates and make the nature of Bell
inequalities easy to understand. Let us consider a particle
with a slippery shape property that is either square or round, depending on
which way we look at it. The particle cannot be seen from two directions at
once, and looking at it changes how it might have looked from other
directions. A source creates entangled pairs of these particles, so that if
we look at the two from the same angle they have the same shape, and sends
them in opposite directions. Shape detectors independent of each other and of
the source are placed in the path of each particle and randomly 
change between three observing angles after the particles are emitted. Because
the particles are entangled, the detectors report the same shape every time
they happen to measure a pair from the same observation angle. Additionally
the detectors measure the same shape for half of all runs when they 
are set arbitrarily and independently to one of the three angles. This last
property does hold for some real systems, and is the key Bell found to show
the existence of alocality. \\In an effort to construct a model for this
situation which is local in nature, we must assume that the information for
shape appearance at each angle is carried on the particles. This is the only
local way to ensure that the same shape is measured every time the detector
angles happen to be the same. We can represent this information 
by either a $s$ (for square) or $r$ (for round) in three slots corresponding to
the three detector angles. Remember that the shape is slippery; we can only
observe the shape from one angle at a time, and subsequent measurement will
not reflect what the shape "would have been" from another angle. Thus we can
learn only two of the three pieces of information by measurement, one from
each particle. The third number in each particle's instruction set is an
unknowable, hidden variable. Suppose a pair of entangled particles which look
square from angles $1$ and $2$ and round from angle $3$ each carry the
instruction set $ssr$. For this particular instruction set, there are five
possible detector settings which yield the same shape $(11, 22, 33, 12, 21)$
and four settings which yield different shapes $(13, 23, 32, 31)$, so the
probability of detecting the same shape given this instruction set is
$\frac{5}{9}$. There are five more possible instruction sets that also give
probability $\frac{5}{9}$ for detecting the same shape. These are $rss, srs,
rrs, rsr$ and $srr$. The only other possible instruction sets are $rrr$ and
$sss$, for which the same shape is measured with probability $1$. Whatever the 
distribution of these instruction sets among the entangled pairs, the
detectors will  measure the same shape in at least $\frac{5}{9}$ of all runs.\\ 
The inequality $P\geq \frac{5}{9}$, where $P$ is the proportion over all runs
that the detectors measure the same shape, is a Bell inequality for this
particular local hidden variable model. However, one of the required features
of any model is that it allows the observed behavior, that the same shape is
observed in only half of all runs. Our inequality is violated;
$P=\frac{1}{2}\not\geq\frac{5}{9}$, so our local hidden variable model does
not adequately describe the system.\\
It is worth noting that this system can be created physically with
spin-entangled electron/positron pairs substituted for the shape-entangled
particles, and Stern-Gerlach analyzers substituted for shape detectors. The
proper three angles to give $P=\frac{1}{2}$ are $0^{\circ}, 120^{\circ}$ and
$240^{\circ}$. The analog for a polarization entangled photon system is polarization detectors at angles $0^{\circ}, 60^{\circ}$ and $120^{\circ}$ but
because the linear polarizers only pass the vertical polarization of its
rotated basis, measurements must be taken at the orthogonal angles as well.

\section{Schr\"odinger's Verschr\"ankung}
Shortly after Bohr's reply to EPR paper on the incompleteness of quantum
theory, Schr\"odinger published a response to EPR in which he introduced the
notion of 'entanglement' to describe such
quantum correlations. In his view that entanglement was the essence of quantum
mechanics and that it illustrates the difference between the quantum and
classical worlds in the most pronounced way. Schr\"odinger realized that the
members of an entangled collection of objects do not have their own individual
quantum states. Only the collection as a whole has a well-defined state.
\\
In quantum mechanics we can prepare two particles in such a way that the
correlations  between them cannot be explained classically, in the sense that
the nature of the correlations we are interested in does not correspond to the
statistics of the particles. Such quantum states are called 'entangled'
states.
\\ 
With the formulation of Bell inequalities and their experimental violation, 
it seemed that the question of non-locality in quantum mechanics had been
settled once for all. In last years the literature has discussed that this
conclusion was a bit to early. Entanglement in mixed quantum states presents
special features not shown when dealing with pure quantum states, to the point
that a mixed quantum state $\rho$ does not violate Bell inequalities, but can
nevertheless reveal quantum mechanical correlations (Werner, 1989).
\\
The motivation of quantum entanglement has following motivations: It plays an
essential role in several counter-intuitive consequences of 
quantum mechanics, is a challenging problem
of quantum mechanics, a role in quantum information theory and quantum
computation. Entanglement, together with quantum parallelism is the
heart of quantum computing.  
In spite of over 100 years of quantum theory with great achievements, we still
know very little about nature.

\section{Erwin Schr\"odinger's ghost cat}
Schr\"odinger (1935) introduced his famous cat.  Schr\"odinger devised his cat
experiment in an 
attempt to illustrate the incompleteness of the theory of quantum mechanics
when going from subatomic to macroscopic systems. Schr\"odinger's legendary
cat was doomed to be killed by an automatic device triggered by the decay of a
radioactive atom. He thought that it could be
both dead and alive. A strange superposition of 

\begin{eqnarray}\label{psiSchr}
|\Psi\rangle \,&=&\,
\frac{1}{\sqrt{2}}\,\big(|\rm{excited\,atom},\rm{alive\,cat}\rangle 
\,\\ &&+\, |\rm{non-excited\,atom},\rm{dead\,cat}\rangle \big)
\end{eqnarray}
was conceived. But the wavefunction (\ref{psiSchr}) showed
no such commitment, superposing  the probabilities. Either the wavefunction
(\ref{psiSchr}), as given by the  Schr\"odinger equation, was not everything,
or it was not right.  
\\
The Schr\"odinger's cat puzzle deals with the superposition principle. If
$|0\rangle$ and $|1\rangle$ are two states, quantum mechanics tells us that
the linear combination $a|0\rangle+b|1\rangle$ is also a possible
state. Whereas such superpositions 
of states have been extensively verified for microscopic systems, the
application of the formalism to macroscopic systems appears to lead
immediately to hard problems in our understanding of the world. Neither has a
book ever observed to be in a superposition of macroscopically distinguishable
positions, nor seems our Schr\"odinger cat that is a superposition of being
alive and dead to bear much resemblance  
to reality as we perceive it. The problem is then how to reconcile the range 
of the Hilbert space of possible states with the observation of a comparably
few ``classical" macroscopic states. 
\\
The long standing puzzle of the Schr\"odinger's cat problem has been largely 
resolved in terms of quantum decoherence.  
\\
The well-known phenomenon of quantum entanglement had  demonstrated that
correlations between systems can lead to counterintuitive properties of the
composite system that cannot be  
composed from the properties of the individual systems. It is the great merit 
of decoherence to have emphasized the ubiquity and essential inescapability of 
system environment correlations.
\\
The Schr\"odinger cat points out the 
paradoxes of playing quantum games with macroscopic objects. For quantum
systems, even at mesoscopic scales, 
decoherence presents a formidable drawback to the maintenance of quantum 
coherence, which is the main drawback in the physical implementation 
of quantum computing. Decoherence typically takes place on extremely 
short time scales. In general, the effect of decoherence 
increases with the size of the system, but it is important to note 
that there exist, admittedly somewhat exotic, examples 
where the decohering influence of the environment can 
be sufficiently shielded as to lead to mesoscopic and even
macroscopic superpositions.

\section{Quantum Information Theory}
The theory of quantum information is a result of the effort
to generalise classical information theory to the quantum world. Quantum
information theory aims to answer the following question:
\begin{quote}   
What happens if information is stored in a state
of a quantum system?
\end{quote}
It is a strength of classical information theory that it does not need to ask
the question about the physical representation of information: There is no
need for a 'ink-on-paper' information theory or a 'DVD information'
theory. This is due to that fact that it is always possible to efficiently
transform information from one representation to another representation. For
this reason, one might be tempted to believe that it is not important whether
information is stored in classical systems or in quantum systems. However this
is not the case: it is not possible, for example, to write down the previously
unknown information contained in the polarisation of a photon of ink on a
paper. In general quantum mechanics does not allow us to read out the state of
an quantum system with arbitrary precision. The existence of Bell
correlations between quantum systems cannot be converted into classical
information. It is possible to transform quantum information
between quantum systems of sufficient quantum information capacity. The
quantum information content of a quantum message $\mc{M}$ can for this reason
be measured in therms of the minimum number $n$ of two-level systems which are
needed to store the message: $\mc{M}$ consists of $n$ qubits. \\
In its original quantum information theoretical sense, the term qubit is thus
a measure for the amount of information. A two-level quantum system can carry
at most one qubit, in the same sense a classical binary digit can carry at
most one classical bit. The term qubit is used as a synonym for a two-level
quantum system. \\
A pure one qubit state is specified by two real parameters, in this sense
quantum information is similar to analog (in contrast to digital) classical
information. Analog information processing seems to be much more efficient
than digital information processing on a first sight, since an analog
information carrier could contain an infinite amount of information. However,
analog information processing is being, or is already been, replaced by
digital information processing. From this one can see, that in practise
analog information processing performs more than digital information
processing. \\
In the presence of noise, which is responsible for this gap between the
theoretical promise and the practical application of analog information. 
In the case of noise, the information content of an analog information
carrier is no longer infinite, but finite. This is a consequence
of Shannon's noisy coding theorem. It is very difficult to protect the
remaining finite information content of analog information carriers against
noise. The example of classical analog information shows that quantum
information processing schemes must necessary be tolerant against noise,
otherwise there would be a chance for them to be useful. It was a big
break through for the theory of quantum information, when quantum error
correction codes and fault-tolerant quantum computation schemes were
discovered.   

\section{Entropy with connection to von Neumann }
John von Neumann  contributed rigorously to establish the correct mathematical
framework for quantum mechanics  with his work {\it Mathematische Grundlagen
  der Quantenmechanik}. He provided in this work a theory of measurement, where the usual notion of wave collapse
is described as an irreversible process (the so called von Neumann or
projective measurement). 
\\
The density matrix was introduced, with different motivations, by
von Neumann and by Landau. The motivation that led Landau was the
impossibility to describe a subsystem of 
a composite quantum system by a state vector. On the other hand, von Neumann
introduced the density matrix in order to develop both quantum statistical
mechanics and a theory of quantum measurements.  
Ideas and methods from information theory are useful in
the study of the probability distributions appearing in quantum mechanics.
Probabilities in quantum mechanics arise in two different ways. On the one hand,
we have the probability distribution
$\tilde p_i = |a_i|^2$, associated with the expansion of a pure quantum state 
$| \Psi \rangle $
in a given orthonormal base $| \psi_i \rangle $,
\begin{eqnarray} \label{qp2}
   | \Psi \rangle \, = \, \sum_i \, a_i \, | \psi_i \rangle,
\end{eqnarray}
where $\sum_i |a_i|^2 = 1$. On the other hand, we have the probabilities $p_i$
appearing when we express the statistical operator $\hat \rho$ as a linear
combination of projector operators, 
\begin{eqnarray} \label{qp1}
\hat \rho \, = \, \sum_i \, p_i | \phi_i \rangle \langle \phi_i |,
\end{eqnarray}
where $\sum_i p_i=1 $, and the states $| \phi_i \rangle$ are not necessarily
orthogonal. Here the statistical operator $\hat \rho$ describes a mixed
quantum state associated with an incoherent mixture of states where each
(pure) state $|\phi_i \rangle $ appears with probability $p_i$. A quantum
mechanical statistical operator differs in fundamental ways from a classical
probability distribution. Nevertheless, the second kind of probability
distributions described above have some similarities with the standard
probability distributions describing classical statistical ensembles
(Sakurai, 1985). On the contrary, the first kind of probabilities, 
associated with pure states, are essentially quantum mechanical in nature and
have no classical counterpart. 
\\
The density matrix formalism was developed to extend the tools of
classical statistical mechanics to the quantum domain. In the classical
framework we compute the partition function of the system in
order to evaluate all possible thermodynamic quantities. 
Von Neumann introduced the density matrix  in a
context of states and operators in a Hilbert space. The knowledge of the
statistical density matrix operator would allow us to compute all average
quantities in a conceptually similar, but mathematically
different way. Let us suppose we have a set of wave functions $|\Psi
\rangle$ which depend parametrically on a set of quantum numbers $n_1, n_2,
..., n_N$. The natural variable which we have is the amplitude with which a
particular wavefunction of the basic set participates in the actual
wavefunction of the system. Let us denote the square of this amplitude by
$p(n_1, n_2, ..., n_N)$. The goal is to make this quantity $p$ to the
classical density function in phase space. We have to verifies that 
$p$ goes over into the density function in the classical limit and that it has
ergodic properties. After checking that $p(n_1, n_2, ..., n_N)$ is a constant
of motion, an ergodic assumption for the probabilities $p(n_1, n_2, ..., n_N)$
makes $p$ a function of the energy only . 
\\
After this procedure, one finally arrives to the density matrix formalism when
seeking a form where $p(n_1, n_2, ..., n_N)$ is invariant with respect to the
representation used. In the form it is written, it will only yield the correct
expectation values for quantities which are diagonal with respect to the
quantum numbers $n_1, n_2, ..., n_N$. \\ 
Expectation values of operators which are not diagonal involve the phases of
the quantum amplitudes. Suppose we subsume the quantum numbers $n_1, n_2, ...,
n_N$ by the single index $i$ or $j$. Then our wave function has the form 

   \begin{equation}
   |\Psi \rangle \,=\,\sum_i a_i\, | \psi_i \rangle. 
   \end{equation}
The expectation value of an operator $B$ which is not 
diagonal in these wave functions, so

   \begin{equation} \label{obs}
   \bbbe(B) \,=\,\sum_{i,j} a_i^{*}a_j\, \langle i| B |j\rangle.
   \end{equation}
The role which was originally reserved for the quantities $|a_i|^2$ is 
thus taken over by the density matrix of your system $S$.

   \begin{equation} \label{mrho}
    \langle j|\,\rho \, |i\rangle \,=\,a_j\, a_i^{*}.
   \end{equation} 
Therefore (\ref{obs}) reads as
   \begin{equation} \label{trc} 
   \bbbe(B) \,=\,\tr (\rho \, B).
   \end{equation}
The invariance of (\ref{trc}) is described by matrix 
theory. We described a mathematical 
framework where the expectation of quantum operators, as described 
by matrices, is obtained by tracing the product of the density operator $\hat
\rho$ times an operator $\hat B$ (Hilbert scalar product between
operators). The matrix formalism here is in the statistical
mechanics framework, although it applies as well for finite quantum systems,
which is usually the case, where the state of the system cannot be described
by a pure state, but as a statistical operator $\hat \rho$ of the form
(\ref{qp1}). Mathematically, $\hat \rho$ is a positive,
semidefinite hermitian matrix with unit trace.
\\  
Given the density matrix $\rho$, von Neumann defined the entropy as 	   
\begin{equation} \label{Sv} 
  S(\rho) \,=\,-{\rm Tr} (\rho \, {\rm ln} \rho),
\end{equation}
which is a proper extension of Shannon's entropy to the quantum case. 
To compute (\ref{Sv}) one has to find a basis in
which $\rho$ possesses a diagonal representation. We note that the entropy
$S(\rho)$ times the Boltzmann constant 
$k_B$ equals the thermodynamical or physical entropy. If
the system is finite (finite dimensional matrix representation) the entropy
(\ref{Sv}) describes the departure of our system from a pure state. In other
words, it measures the degree of mixture of our state describing a given
finite system. Properties of the von Neumann entropy 
\begin{enumerate}
\item  $S(\rho)$ is only zero for pure states. 
\item $S(\rho)$ is
  maximal and equal to ln$N$ for a maximally mixed state, $N$ being the
  dimension of the Hilbert space.
\item $S(\rho)$ is invariant under changes in the basis of $\rho$, that is, 
$S(\rho)=S(U\,\rho \, U^{\dagger})$, with $U$ a unitary transformation. 
\item $S(\rho)$ is concave, that is, given a collection of positive numbers
  $\lambda_i$ and density operators $\rho_i$, we have

\begin{equation} 
  S\bigg(\sum_{i=1}^k \lambda_i \, \rho_i \bigg) \,\geq\, \sum_{i=1}^k
  \lambda_i \, S(\rho_i). 
\end{equation}

\item $S(\rho)$ is additive. 
  \\Given two density matrices $\rho_A,\rho_B$
  describing different systems $A$ and $B$, then $S(\rho_A \otimes
  \rho_B)=S(\rho_A)+S(\rho_B)$. Instead, if $\rho_A,\rho_B$ are the reduced
  density matrices of the general state $\rho_{AB}$, then

\begin{equation} 
 |S(\rho_A)\,-\,S(\rho_B)|\,\leq \, S(\rho_{AB}) \, \leq \,
 S(\rho_A)\,+\,S(\rho_B). 
\end{equation}

This property is known as subadditivity. While in Shannon's theory 
the entropy of a composite system can never be lower than the entropy of any
of its parts, in quantum theory this is not the case. Actually, this can be
seen as an indicator of an entangled state $\rho_{AB}$.

\item The von Neumann entropy (\ref{Sv}) is strongly subadditive:

\begin{equation} 
S(\rho_{ABC}) \, + \, S(\rho_{B}) \, \leq \, S(\rho_{AB}) \,+\, S(\rho_{BC}).
\end{equation}

\end{enumerate} 
The von Neumann entropy is being extensively used in different forms
(conditional entropies, relative entropies, etc.) in all the framework of
quantum information theory. Entanglement measures are based upon some
quantity 
directly related to the von Neumann entropy. However, there have appeared in
the literature  several papers dealing with the possible 
inadequacy of the Shannon information, and consequently of the von Neumann 
entropy as an appropriate quantum generalization of Shannon entropy. The main 
argument is that in classical measurement the Shannon information is a natural 
measure of our ignorance about the properties of a system, whose existence is 
independent of measurement. Conversely, quantum measurement cannot be claimed
to reveal the properties of a system that existed before the measurement was
made. This controversy have encouraged some authors  to introduce the 
non-additivity property of Tsallis' entropy as the main reason for recovering 
a true quantal information measure in the quantum context, claiming that
non-local correlations ought to be described because of the particularity of
Tsallis' entropy.

\end{document}